\documentclass{ws-ijait}
\PassOptionsToPackage{hyphens}{url}
\usepackage{listings}

\usepackage{pgfplots}
\usepgfplotslibrary{units}
\pgfplotsset{compat=newest} 

\usepackage{tikz}
\usetikzlibrary{patterns}
\usepackage{algorithm2e}

\begin{document}

\markboth{C. S. Kouzinopoulos, J.-A. M. Assael, Th. K. Pyrgiotis, K. G. Margaritis}
{A Hybrid Parallel Implementation of the Aho-Corasick and Wu-Manber Algorithms Using CUDA and MPI}

%
\catchline{}{}{}{}{}
%

\title{A Hybrid Parallel Implementation of the Aho-Corasick\\
and Wu-Manber Algorithms Using NVIDIA CUDA and MPI\\
Evaluated on a Biological Sequence Database}

\author{Charalampos S. Kouzinopoulos, John-Alexander M. Assael, Themistoklis K. Pyrgiotis, and
Konstantinos G. Margaritis}

\address{Parallel and Distributed Processing Laboratory\\
Department of Applied Informatics, University of Macedonia\\
156 Egnatia str., P.O. Box 1591, 54006 Thessaloniki, Greece\\
charalampos.kouzinopoulos@cern.ch, john.assael@wolfson.ox.ac.uk, t.pirgiot@gmail.com, kmarg@uom.gr}

\maketitle

\begin{history}
\received{(Day Month Year)}
\revised{(Day Month Year)}
\accepted{(Day Month Year)}
\end{history}

\begin{abstract}Multiple matching algorithms are used to locate the occurrences of patterns from a finite pattern set in a large input string. Aho-Corasick and Wu-Manber, two of the most well known algorithms for multiple matching require an increased computing power, particularly in cases where large-size datasets must be processed, as is common in computational biology applications. Over the past years, Graphics Processing Units (GPUs) have evolved to powerful parallel processors outperforming Central Processing Units (CPUs) in scientific calculations. Moreover, multiple GPUs can be used in parallel, forming hybrid computer cluster configurations to achieve an even higher processing throughput. This paper evaluates the speedup of the parallel implementation of the Aho-Corasick and Wu-Manber algorithms on a hybrid GPU cluster, when used to process a snapshot of the Expressed Sequence Tags of the human genome and for different problem parameters.\end{abstract}


\section{Introduction}
\label{sec:introduction}
The endless market demand for better and more realistic computer graphics evolved Graphics Processing Units into powerful and highly parallel multicore processors with enormous computational power. Moreover, their parallel nature facilitates the rapid execution of scientific calculations, outperforming in many cases traditional CPUs. Nowadays, various APIs have been introduced to enable the development and execution of General Purpose Computations on a GPU \cite{gpgpuorg} (GPGPU). One of the most widely known APIs for GPGPU is NVIDIA CUDA.\cite{CUDA_SDK}

Multiple pattern matching is a variant of the string matching problem. It is used to locate all the positions in an input string where one or more patterns from a finite pattern set occur. Computational Biology is a major area where the multiple pattern matching problem is applicable since many tasks require the location of nucleotide or Amino Acid sequence patterns in biological sequence databases. The multiple pattern matching problem can be defined as,\cite{Kouzinopoulos2011}:

\textit {\textbf{Definition}. Given an input string $T = t_{0}t_{1}\ldots t_{n-1}$ of size $n$ and a finite set of $d$ patterns $P = {p^{0}, p^{1},\ldots, p^{d-1}}$, where each $p^{r}$ is a string $p^{r} = p^{r}_{0} p^{r}_{1}\ldots p^{r}_{m-1}$ of size $m$ over a finite character set $\Sigma$, the alphabet size is denoted as $|\Sigma|$ and the total size of all patterns as $|P|$, the task is to find all occurrences of any of the patterns in the input string. More formally, for each $p^r$ find all $i$ where $0 \leq i < n - m + 1$ such that for all $j$ where $0 \leq j < m$ it holds that $t_i = p^r_j$}

This paper presents a hybrid implementation of the Aho-Corasick \cite{Aho1975} (AC) and Wu-Manber \cite{Wu1994} (WM) multiple pattern matching algorithms on an MPI cluster using the CUDA API. The performance of both the sequential and the parallel implementations of the algorithm was evaluated on a homogeneous cluster of nodes for various problem parameters, including different pattern set and cluster node sizes. The data set used for the experiments of this paper consisted of a snapshot of the Expressed Sequence Tags (ESTs) and different sets of patterns.

The paper is organized as follows: Related work is presented in section \ref{sec:related-work}. Section \ref{sec:background} details the way the algorithms work. The GPU and MPI architectures and implementations are introduced in section \ref{sec:distrcudawm}. Experimental methodology and results are given in sections \ref{sec:experimental-methodology} and \ref{sec:experimental-results} respectively. Finally, the conclusions of this research are presented in section \ref{sec:conclusions}.

\section{Related Work}
\label{sec:related-work}

Several implementations of multiple pattern matching algorithms running on GPUs have been introduced during the last years, offering a substantial performance increase compared to their sequential versions. Some major fields of interest on these implementations include bioinformatics and intrusion detection systems.

In 2008, a group of researchers proposed a GPU version of the Wu-Manber multiple pattern matching algorithm.\cite{Huang2008} The algorithm was to be used in network intrusion detection, and was implemented under OpenGL in order to take the advantage of an NVIDIA GeForce 7600 GT card. The experiments proved to be two times faster than the existing optimized version that was used in Snort.\cite{SnortWeb}

Later in 2011,\cite{Hongjian2011} a group of researchers proposed an optimized version of the agrep algorithm,\cite{Wu1992} which is based on Wu-Manber using the CUDA API and taking advantage of a GeForce GTX285, for approximate nucleotide sequence matching. The performance of the implementation was evaluated for sequences of genomes, comparing an OpenMP implementation to the CUDA implementation of the algorithm, and proved to exhibit $70$-fold and $36$-fold performance speedups, for pattern sizes of $30$ and $60$ respectively.

Moreover, a modified version of the Wu-Manber algorithm for approximate matching was presented in 2011.\cite{tran2011} The implementation was simplified to run on a NVIDIA GeForce 480 using the OpenCL API, and managed to achieve $62$-fold speedups. The next year, another group of researchers,\cite{pyrgiotis2012parallel} implemented the Wu-Manber algorithm using OpenCL and conducted an extended research and analysis on texts with different alphabet sizes, including biological sequences, and proved to have a $31$x speedup, running the experiments on a NVIDIA GTX 280.

Finally in 2012, distributed hybrid implementations of several multiple pattern matching algorithms were presented,\cite{Kouzinopoulos2012} using MPI. One of the algorithms was Wu-Manber, and reached a maximum of $19.2$x performance speedup. The comparison was between the single-threaded sequential CPU implementation and a multithreaded implementation running on $10$ dual-core nodes.

The Aho-Corasick algorithm was implemented in \cite{Vasiliadis2008} using the CUDA API to perform network intrusion detection. The Aho-Corasick trie was represented by a two-dimensional \textit{state transition} array; each row of the array corresponded to a state of the trie, each column to a different character of the alphabet $\Sigma$ while the cells of the array represented the \textit{next} state. The array was precomputed by the CPU and was then stored to the texture memory of the GPU. The input string (in the form of network packets) was stored in buffers allocated using \textit{pinned} host memory using a double buffering scheme and was copied to the global memory of the GPU as soon as each buffer was full. Since the input string was in the form of network packets, two different parallelization approaches were considered; with the first, each packet was processed by a different warp of threads while with the second, each packet was processed by a different thread. A similar implementation of the Aho-Corasick algorithm was used in \cite{Vasiliadis2010} to perform heavy-duty anti-malware operations.

The Parallel Failureless Aho-Corasick algorithm (PFAC), an interesting variant of the Aho-Corasick algorithm on a GPU was presented in \cite{Lin2010}. It is significantly different than the rest of the parallel implementations of the same algorithm in the sense that each character of the input stream was assigned to a different thread of the GPU. Since each thread needs only to examine if a pattern exists \textit{starting} at the specific character and no back-tracking takes place, the \textit{supply} function of Aho-Corasick was removed. The \textit{goto} function was mapped into a two-dimensional \textit{state transition} array. The \textit{state transition} array was then stored in shared memory by grouping the patterns based on their prefixes and distributing these groups into different multiprocessors.

In \cite{Tumeo2011}, the Aho-Corasick algorithm was implemented using the CUDA API. The Aho-Corasick trie was represented using a two-dimensional \textit{state transition} array that was precomputed on the CPU and stored to the texture memory of the GPU. Instead of storing a map of the final states in another array, the final states that corresponded to a complete pattern were flagged directly in the \textit{goto} array by reserving $1$ bit. Bitwise operations were then used to check its value. The parallelization of the algorithm was achieved by assigning different characters of the input string to different threads of the GPU and letting them perform the matching by accessing the shared \textit{state transition} array.

The work presented in \cite{Zha2011} focused on the implementation of the Aho-Corasick algorithm on a GPU. The Aho-Corasick trie was precomputed on the CPU and was stored in the texture memory of the GPU but it is not clear the way it was represented. The input string was stored in the global memory of the GPU, partitioned to blocks and assigned to different threads in order to achieve parallelization. The threads were then responsible to process the input string and create an output array of the states of the Aho-Corasick trie that corresponded to each character position. The paper utilized a number of optimizations in order to further improve the performance of the algorithm implementation; casting the input string from \textit{unsigned char} to \textit{uint4} to ensure that each thread will read $16$ input string characters from the global memory instead of $1$. To further improve the bandwidth utilization, the accesses of multiple threads inside the same half-warp were coalesced by reading the required data to process an input string block to the shared memory of the device. Finally, to avoid shared memory bank conflicts, the threads of a half-warp were accessing memory from different banks.

An implementation of the Aho-Corasick algorithm was also presented in \cite{Hu2012} using the CUDA API. Similar to the methodology used in previously published research papers, the trie of the algorithm was represented using a two-dimensional \textit{state transition} array and was stored compressed in the texture memory of the GPU while the input string was stored in global memory. Kargus was introduced in \cite{Jamshed2012}, a software that utilizes the Aho-Corasick algorithm to perform intrusion detection on a hybrid system consisting of multi-core CPUs and heterogeneous GPUs. The trie of Aho-Corasick was represented in the GPU using a two-dimensional \textit{state transition} array. The \textit{state transition} array was created in the CPU and was stored in the GPU. The network packets that comprise the input string were stored in texture memory. To increase the utilization of the memory bandwidth of the GPU, the input string was cast to \textit{uint4} using a technique similar to \cite{Zha2011}.

In \cite{Pungila2012}, the Aho-Corasick and Commentz-Walter algorithms were used to perform virus scanning accelerated through a GPU. The Aho-Corasick and Commentz-Walter tries were represented in the GPU using stacks. The \textit{goto} and \textit{supply functions} were substituted with offsets, essentially serializing the trie in a continuous memory block. The stacks were precomputed in the CPU and were then transferred to the GPU. Parallelization was achieved using a data-parallel approach.

Finally, the Aho-Corasick algorithm was implemented in \cite{Tumeo2012} for a heterogeneous computer cluster with $5$ NVIDIA Tesla S1070 boxes, each box being the equivalent of $4$ C1060 GPUs, for a total of $20$ GPUs.

To the best of our knowledge, this is the first time that the Aho-Corasick and Wu-Manber algorithms are implemented in parallel on a hybrid CUDA/MPI parallel cluster architecture.

\section{Background}
\label{sec:background}

\subsection{Aho-Corasick}

Aho-Corasick is an extension of the Knuth-Morris-Pratt algorithm for a set of patterns $P$. It uses a deterministic finite state pattern matching machine; a rooted directed tree or \textit{trie} of $P$ with a \textit{goto} function $g$ and an additional \textit{supply} function $Supply$. The \textit{goto} function maps a pair consisting of an existing state $q$ and a symbol character into the next state. It is a generalization of the \textit{next} table or the \textit{success} link of the Knuth-Morris-Pratt algorithm for a set of patterns where a parent state can lead to its child states by $\sigma$ where $\sigma$ is a matching character. Each state of the trie is labeled after a single character of a pattern $p^r \in P$. If $L(q)$ denotes the label of the path between the initial state and a state $q$, then $L(q)$ is also a prefix of one of the patterns. For each pattern $p^r$ there is a state $q$ such that $L(q) = p^r$. This state is marked as terminal and when visited during the search phase indicates that a complete match of $p^r$ was found. The \textit{supply} function of Aho-Corasick is based on the \textit{supply} function of the Knuth-Morris-Pratt algorithm. It is used to visit a previous state of the automaton when there is no transition from the current state to a child state via the \textit{goto} function.

\begin{figure}[h]
\centering
{\epsfig{file = 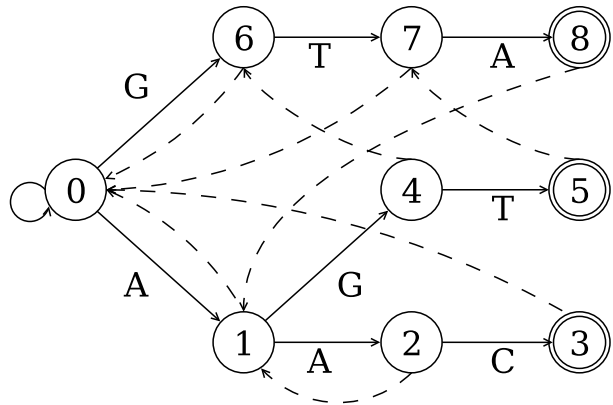, width = 8.0cm}}
  \caption[The automaton of the Aho-Corasick algorithm]{The automaton of the Aho-Corasick algorithm for the pattern set ``AAC", ``AGT" and ``GTA" }
  \label{fig:ACtransitiondiagram}
\end{figure}

The \textit{goto} function and the \textit{supply} function are constructed during the preprocessing phase. To build the \textit{goto} function, the trie is depth-first traversed and extended for each character of the patterns from a finite pattern set $P$ at the same time the outgoing transitions to each state are created. The \textit{supply} function is built in transversal order from the trie until it has been computed for all states. For each state $q$, the \textit{supply} link can be determined based on the longest suffix of $L(q)$ that is also a prefix of any pattern from $P$. Assume that for the parent state $q_{parent}$ of $q$, $g(q_{parent}, \sigma) = q$. If $Supply(q_{parent})$ also has an outgoing transition to a state $h$ by $\sigma$, then the \textit{supply} state of $q$ can be set to $h$. In any other case, $Supply(Supply(q_{parent}))$ must be checked for a transition to a state by $\sigma$ and so on, until one such state is found or it is determined that no such state exists; in that case, the \textit{supply} state of $q$ is set to the initial state.

\begin{algorithm}[H]

Function AC\_Preproc\_Goto ( $p, m, d, \Sigma$ )\\

create state $q_0$

\ForAll{ $\alpha \in \Sigma$ } {

	$g(q_0, \alpha) := fail$
}

\For{ $i := 0; i < d; i := i + 1$ } {

	$j := 0$; $state := q_0$\\
	
	\While{ $newState := g(state, p^i_j) \neq fail$ }{
	
		$state := newState$; $j := j + 1$\\
	}
	
	\For{ $l := j; l < m; l := l + 1$} {
		
		create state $q_{current}$\\
		
		\ForAll{ $\alpha \in \Sigma$ } {

			$g(q_{current}, \alpha) := fail$

		}
		
		$newState := q_{current}$\\
	
		$g(state, p^i_l) := newState$\\
	
		$state := newState$\\
	}
	
	$Output(q_{current}) := Output(q_{current}) \cup \{p^i\}$\\
	Add terminal state on $q_{current}$\\
}

\caption{The construction of the \textit{goto} function $g$ of the Aho-Corasick automaton}
\label{compl:aho_corasick_preprocessing_goto}
\end{algorithm}

Let $u$ be the longest suffix of the input string $t_0\ldots t_{i-1}$ that is also a prefix of any pattern $\in P$. The character $\sigma$ located at position $i$ of the input string is scanned next. If there is an outgoing transition from the current state $q$ to another state $f$ as indicated by the \textit{goto} function, then $L(f) = u\sigma$ is the new longest suffix of the input string at position $i$ that is a prefix of one of the patterns. A match of a pattern exists in the input string if $|u\sigma| = m$. If, on the other hand $g(q, \sigma) = fail$, then $g(Supply(q), \sigma)$ is checked for an outgoing transition by $\sigma$. If $g(Supply(q), \sigma)$ leads to a state $f'$ then $u = L(f')$. If $g(Supply(q), \sigma) = fail$ then $g(Supply(Supply(q)), \sigma)$ is considered and so on, until an outgoing transition by $\sigma$ is found or until the supply state of the initial state is reached; in that case, the search will start again from the initial state. The construction of the \textit{goto} function of the Aho-Corasick automaton is given in Algorithm~\ref{compl:aho_corasick_preprocessing_goto}, the computation of the \textit{supply} function is presented in Algorithm~\ref{compl:aho_corasick_preprocessing_supply} while the search phase of the Aho-Corasick algorithm is given in Algorithm~\ref{compl:aho_corasick_search}. The output function returns $L(q)$ for each terminal state $q$ and is denoted as $Output()$. A transition that does not point to a state is denoted as $fail$.

\begin{algorithm}[H]

Function AC\_Preproc\_Supply ( $\Sigma$ )\\

\ForAll{ $\alpha \in \Sigma$ } {

	\eIf{$g(q_0, \alpha) = fail$} {
		$g(q_0, \alpha) := q_0$\\
	}{
	
		$Supply(g(q_0, \alpha)) := q_0$\\
	
	}
}

\ForAll{ $currentState \in$ trie states in transversal order } {

	\ForAll{ $\alpha \in \Sigma$ } {

		$s := g(currentState, \alpha)$\\
		
		\If{$s \neq fail$} {

			$state := Supply(currentState)$\\
		
			\While{ $g(state, \alpha) = fail$}{
	
				$state := Supply(state)$\\
			}
			
			$Supply(s) := g(state, \alpha)$\\
		}
	}
}

\caption{The construction of the \textit{supply} function $Supply$ of the Aho-Corasick automaton}
\label{compl:aho_corasick_preprocessing_supply}
\end{algorithm}

The \textit{goto} function can be implemented using any of the following data structures: an array of size $|\Sigma|$, where each state has an outgoing transition for every character of the alphabet by precomputing all the transitions simulated by the \textit{supply} function \cite{Navarro2002}; a linked list that is space efficient but not time efficient; or a balanced search tree that is considered as a heavy-duty compromise and often not practical \cite{Dori2006}. The implementation used for the experiments of this paper was based on code from the Streamline system I/O software layer \cite{WEB03}. It uses a linked list for the \textit{supply} function and a linked list of arrays to represent the transitions of the \textit{goto} function with each cell of the arrays potentially containing a pointer to the next node. Each list node corresponds to a different state of the trie and has an array of size $|\Sigma|$ with an outgoing transition for every character of $\Sigma$. The trie of $P$ can then be built for all $m$ patterns in $\mathcal{O}(|\Sigma|m^2)$ time, with a total size of $\mathcal{O}(|\Sigma|m^2)$. The time to pass through a transition of the \textit{goto} function is $\mathcal{O}(1)$ in the worst and average case, while the search phase has a cost of $\mathcal{O}(n)$ in the worst and average case.

\begin{algorithm}[H]

Function AC\_Search ( $t, m, n$ )\\

$state := q_0$\\
\For{ $i := 0; i < n; i := i + 1$} {
	
	\While{ $newState := g(state, t_i) = fail$ }{
			
		$state := Supply(state)$\\
	}
	
	$state := newState$\\
	
	\If{ $Output(state)$ is not empty }{
	
		report match at $i - m + 1$\\
	
	}
}

\caption{The search phase of the Aho-Corasick automaton}
\label{compl:aho_corasick_search}
\end{algorithm}

An example of a complete Aho-Corasick automaton for the pattern set ``AAC", ``AGT" and ``GTA" is presented in Figure~\ref{fig:ACtransitiondiagram}. Assume that the \textit{goto} function of the trie is already constructed and that the \textit{supply} function for states $0 - 4$ has been computed. The \textit{supply} state of state $5$ is determined next. State $4$ is the parent state of state $5$ since $g(4,``T") = 5$ and $Supply(4) = 6$, therefore the \textit{goto} function of state $6$ is considered next. Since $g(6,``T") = 7$ then $Supply(5)$ can be set to $7$. If there was no outgoing transition from state $6$ by $``T"$ then $Supply(6)$ would be checked next for an outgoing transition to another state by $``T"$ and so on, until one such state is found or is determined that no such state exists.

\subsection{Wu-Manber}

Wu-Manber is a generalization of the Horspool algorithm for multiple pattern matching. It scans the characters of the input string backwards for the occurrences of the patterns, shifting the search window to the right when a mismatch or a complete match occurs. To perform the shift, the \textit{bad character} shift function of the Horspool algorithm is used. The \textit{bad character} shift for a character $\sigma$ determines the safe number of shifts based on the position of the rightmost occurrence of $\sigma$ in \textit{any} pattern. The probability of $\sigma$ existing in one of the patterns increases with the size of the pattern set and is inversely proportional to the alphabet size and thus the maximum possible shift is decreased. To improve the efficiency of the algorithm, Wu-Manber considers the characters of the patterns and the input string as blocks of size $B$ instead of single characters, essentially enlarging the alphabet size to $|\Sigma|^B$.

\begin{figure}[h]
\centering
{\epsfig{file = 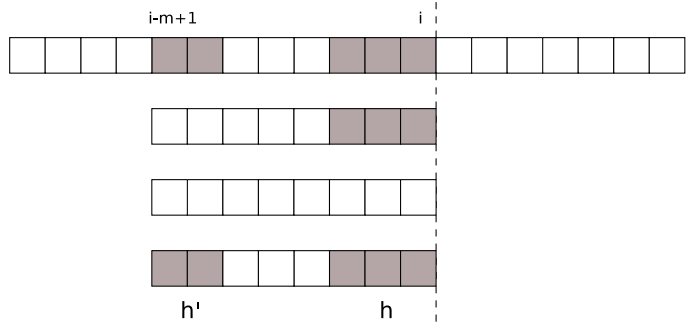, width = 9.0cm}}
  \caption{Comparing the suffix and prefix of the search window of the Wu-Manber algorithm}
  \label{fig:wu_manber_border}
\end{figure}

During the preprocessing phase, three tables are built from the patterns, the \textit{SHIFT}, \textit{HASH} and \textit{PREFIX} tables. \textit{SHIFT} is the equivalent of the \textit{bad character} shift of the Horspool algorithm for blocks of characters, generalized for multiple patterns. If $B$ does not appear in \textit{any} pattern, the search window can be safely shifted by $m - B + 1$ positions to the right. Let $h$ be the hash value of a block of $B$ characters as determined by a hash function $h_1()$. Then, \textit{SHIFT[h]} is the distance of the rightmost occurrence of $B$ to the end of \textit{any} pattern. The \textit{HASH} and \textit{PREFIX} tables are only used when the shift value stored in \textit{SHIFT[h]} is equal to $0$. \textit{HASH[h]} contains an ordered list of pattern indices whose $B$-character suffix has a hash value of $h$. For each of these patterns, let $h'$ be the hash value of their $B'$-character prefix as determined by a hash function $h_2()$. The hash value $h'$ for each pattern $p$ is stored in \textit{PREFIX[p]}. That way, a potential match of the $B$-character suffix of a pattern can be verified first with the $B'$-character prefix of the pattern before comparing the patterns directly with the input string. The complexity of Wu-Manber was not given in the original paper, since hash functions $h_1()$ and $h_2()$ were not specified and the size of the \textit{SHIFT}, \textit{HASH} and \textit{PREFIX} tables was not given.\cite{Navarro2002} For the experiments of this paper, the algorithm was implemented with a block size of $B = 3$ and $B' = 2$ while hash values $h$ and $h'$ were calculated by shift and add; shifting the hash value to the left by \textit{bitshift} positions and then adding in the ASCII value of a pattern or input string character. The value of \textit{bitshift} was set to $2$. Finally, the verification of the patterns to the input string was performed using the \textit{memcmp()} function of \textit{string.h}.

\begin{algorithm}[h]

Function WM\_Preproc ( $p, m, d, B, B'$ )\\

Initialize all elements of $SHIFT$ to $m - B + 1$\\

\For{ $i := 0; i < d; i := i + 1$} {

	\For{ $q := m; q \geq B; q := q - 1$} {
		
		$h := h_1(p^i_{q - B - 1}\ldots p^i_{q - 1})$\\
		
		$shiftlen := m - q$\\
			
		$SHIFT[h] := MIN( SHIFT[h], shiftlen )$\\
			
		\If{ $shiftlen = 0$ }{
			
			$h' := h_2(p^i_0\ldots p^i_{B' - 1})$\\
			
			$HASH[h] := HASH[h] \cup \{i\}$\\
			
			$PREFIX[i] := h'$\\
		}
	}
}

\caption{The preprocessing phase of the Wu-Manber algorithm}
\label{compl:wu_manber_preproc}
\end{algorithm}

Assume that the search window is aligned with the input string at position $i$ and that $h$ is the hash value of the $B$-character suffix of $t_0\ldots t_i$. Then the \textit{SHIFT} table is used to determine the number of safe shift positions. If \textit{SHIFT[h]} $> 0$ then the search window is shifted by \textit{SHIFT[h]} positions. If, on the other hand, \textit{SHIFT[h]} $= 0$, the suffix of the input string potentially matches the suffix of \textit{some} patterns of the pattern set and thus it must be determined if a complete match occurs at that position. The hash value $h'$ of the $B'$-character prefix of the input string starting at position $i - m + 1$ is then computed. For each pattern $p^r$ with the same hash value $h$ of its $B$-character suffix, it is checked if \textit{PREFIX[p]} matches with $h'$. If both the prefix and the suffix of the search window match with the prefix and suffix of some $p^r \in P$, then the corresponding patterns are compared directly with the input string. The preprocessing phase of the Wu-Manber algorithm is detailed in Algorithm~\ref{compl:wu_manber_preproc} while the search phase is presented in Algorithm~\ref{compl:wu_manber_search}.

\begin{algorithm}[h]

Function WM\_Search ( $p, t, m, n, B, B'$ )\\

$i = m - 1$\\

\While { $i < n$ } {

	$h := h_1(t_{i - B}\ldots t_i)$\\

	\uIf { $SHIFT[h] > 0$ } {
		$i := i + SHIFT[h]$\\
	}
	\Else {

		$h' := h_2(t_{i - m + 1}\ldots t_{i - m + 1 + B' - 1})$\\
		
		\ForAll{ pattern indices $r$ stored in $HASH[h]$ } {
		
			\If{ $PREFIX[r] = h'$ } {
		
				Verify the pattern corresponding to $r$ directly against the input string\\
			}
		}
		
		$i := i + 1$\\
	}
}

\caption{The search phase of the Wu-Manber algorithm}
\label{compl:wu_manber_search}
\end{algorithm}

The complexity of Wu-Manber was not given in the original paper, since hash functions $h_1()$ and $h_2()$ were not specified and the size of the \textit{SHIFT}, \textit{HASH} and \textit{PREFIX} tables was not given.\cite{Navarro2002} For the experiments of this paper, the algorithm was implemented with a block size of $B = 3$ and $B' = 2$ while hash values $h$ and $h'$ were calculated by shift and add; shifting the hash value to the left by \textit{bitshift} positions and then adding in the ASCII value of a pattern or input string character. The value of \textit{bitshift} was set to $2$. Finally, the verification of the patterns to the input string was performed using the \textit{memcmp()} function of \textit{string.h}. The cost of the implementation used in the experiments of this paper is as follows. To calculate the values of the \textit{SHIFT}, \textit{HASH} and \textit{PREFIX} tables during the preprocessing phase, the algorithm requires an $\mathcal{O}(|P|)$ time. The space of Wu-Manber depends on the size of \textit{SHIFT}, \textit{HASH} and \textit{PREFIX}. The space needed for the \textit{SHIFT} table is $\displaystyle\sum\limits_{i=0}^{B-1} |\Sigma| \times (2 ^ {bitshift})^i$. In the worst case there could be $d$ patterns with the same hash value $h$ or $h'$ for their $B$-character suffix or $B'$-character prefix respectively, therefore \textit{HASH} and \textit{PREFIX} require a $d \times \displaystyle\sum\limits_{i=0}^{B-1} |\Sigma| \times (2 ^ {bitshift})^i$ space for a space complexity of $\mathcal{O}(d \times \displaystyle\sum\limits_{i=0}^{B-1} |\Sigma| \times (2 ^ {bitshift})^i)$.

In the worst case for the searching phase of the Wu-Manber algorithm, the input string and $m-1$ characters of all $d$ patterns consist of the same repeating character $\sigma$ with the character at position $m - B - 1$ of each pattern being different. The algorithm will then encounter a potential match on every position of the input string since \textit{SHIFT[h]} will constantly be $0$. Therefore, as hash values $h$ and $h'$ of the patterns will be identical, the $m - B$ characters of all $d$ patterns will be compared directly with the input string using the \textit{memcmp()} function. The worst case searching time of Wu-Manber is given in \cite{Chen2005} as $\mathcal{O}(n\log_{|\Sigma|}(|P|)d(m-1))$. In \cite{Navarro2004} the lower bound for the average time complexity of exact multiple pattern matching algorithms is given as $\Omega(n\log_{|\Sigma|}(|P|)/m)$ and according to \cite{Chen2005} the searching phase of the Wu-Manber algorithm is optimal in the average case for a time complexity of $\mathcal{O}(n\log_{|\Sigma|}(|P|)/m)$. In \cite{Liu2005} the average time complexity of Wu-Manber was also estimated as $\mathcal{O}(\frac{n}{(m-B+1)\times(1-\frac{(m-B+1)\times d}{2\times |\Sigma|^B})})$.

\section{Distributed CUDA Implementation}
\label{sec:distrcudawm}

\subsection{GPU Architecture}
\label{sec:GPU-Architecture}

A GPU is a hardware device that acts as a separate co-processor to the host. It is based on a scalable array of multithreaded \textit{streaming multiprocessors} (SMs) that have a different design than CPU cores; they target lower clock rates; they support instruction-level parallelism but not branch prediction or speculative execution; and they have smaller caches \cite{Wilt2013}. Each SM consists of a number of stream processors (SPs), special function units for floating-point arithmetic operations and mathematical functions, one or more warp schedulers and an independent instruction decoder so they can run different instructions. The SPs, or CUDA Cores, are lightweight in-order processors that are used for arithmetic operations. Moreover, each SM has a number of resources including a $32$-bit register file and a shared memory area that as detailed later in this section are distributed among the available threads and thread blocks respectively. On compute capability 1.x GPUs, SMs are grouped into \textit{Texture Processor Clusters} (TPCs) that contain additionally a texture unit and a texture memory cache. On compute capability 2.x and newer GPUs, the SMs are grouped into \textit{Graphics Processing Clusters} (GPCs).

The threads of a CUDA GPU are very lightweight comparing to threads of multicore CPU systems with a very small creation overhead involved. Although they are lightweight in the sense that they operate on small pieces of data, they are fully fledged in the conventional sense, each thread with its own stack, register file, program counter and local memory \cite{Halfhill2008}. The threads, that are identifiable through a unique thread ID using the \textit{threadIdx} variable are organized into \textit{thread blocks} and are then assigned to SMs, with each SM being capable of executing multiple thread blocks. The SMs divide the thread blocks into warps of threads that are queued for work. A warp typically consists of $32$ threads, a half-warp consists of $16$ threads while a quarter-warp consists of $8$ threads. Threads are grouped into warps in a deterministic way; for warps with a size of $32$ threads, the first warp will always contain threads with a thread ID between $0-31$. Stream processors follow a MIMD model for warps since different warps can have different execution paths and process different data. Threads within the same warp though, follow a SIMD or as called by NVIDIA, Single Instruction, Multiple Thread (SIMT) model. The instruction decoder of an SM fetches a common, single instruction that will be executed by all the SPs at the same time, forming the basis of SIMT execution \cite{Lakshminarayana2010}. One instruction is fetched every fetch cycle per warp scheduler for a warp that is selected in a round robin fashion from among the active, ready warps of the SM. The instructions are then placed in a common issue queue from where they are dispatched for execution by the instruction dispatch units. The threads of a warp start at the same program address and execute concurrently a common instruction at every cycle. Individual threads can branch out and execute independently but this comes with a performance penalty. If threads of a warp diverge via a data-dependent conditional branch, such as an \textit{if-else} statement, the warp executes sequentially each branch path taken, disabling threads that are not on that path to ensure the correctness of the results \cite{CUDA_SDK}. The compiler inserts the reconvergence point using a field in the instruction encoding and also inserts an instruction before a diverging branch that provides the hardware with the location of the reconvergence point \cite{Papadopoulou2009}. Using this information, the threads converge back to the same execution path when all paths are complete. 

There are two sources of latency that affect the performance of a kernel, instruction latency and memory latency. Instruction latency is the number of cycles between issuing an arithmetic instruction and the processing of the instruction by the arithmetic pipeline of the GPU. Memory latency is the number of cycles needed to access a memory address. To hide both arithmetic and memory latency, there should be a number of warps \textit{resident} or in other words maintained on-chip during the entire lifetime of the warp on an SM, so that the SM can choose between them instructions to issue. The number of resident warps required to completely hide latency depends on the architecture of the device. A warp may not be ready to execute due to different factors; waiting on register dependencies; accessing off-chip memory; waiting on some synchronization point; or waiting to finish executing the previous instruction. At every instruction issue time, a \textit{warp scheduler} switches from one warp to another and switches contexts between threads. Because the execution context, including program counters and registers, for each warp processed by a multiprocessor is resident on the SMs, context switching is very fast. The resources of an SM are limited, therefore the \textit{occupancy}, or the number of threads and blocks that an SM can maintain, can be affected by different factors; the block size, the shared memory usage and the register usage. An SM can only support a few concurrent resident thread blocks (usually $8$ or $16$), therefore it is important to use blocks with a sufficiently large block size in threads. The amount of shared memory per block and the number of registers per thread used by a kernel also affect significantly the occupancy of an SM. To ensure maximum occupancy, a kernel should use up to

\begin{equation}
\frac{\textrm{Total amount of shared memory per block}}{\textrm{Maximum number of active thread blocks per SM}}
\end{equation}

\noindent
shared memory and 

\begin{equation}
\frac{\textrm{Total number of registers available per block}}{\textrm{Maximum number of threads per SM}}
\end{equation}

\noindent
registers per thread, although as detailed in \cite{Volkov2010}, maximizing occupancy does not always result in a better performance.

GPUs have different memories, both on-chip and off-chip, each with its own advantages and limitations. Unlike the memory architecture of traditional computer systems where the compiler is mainly responsible for distributing data between an off-chip RAM and different levels of cache, the programmer of a GPU application must, in most cases, explicitly copy data between the memory areas of the device, trying to find a balance between the size of each memory area and the cost to access it in terms of clock cycles. Although this model offers the potential for high performance gains, it also comes with an increased coding complexity.

The fastest and at the same time more limited memory area of a GPU is the register file, a highly banked array of processor registers built out of dense SRAM arrays \cite{Gebhart2012}. As already discussed, each SM has its own register file. The threads of each SM use dedicated registers to store their register context in order to perform context switching very fast. To accommodate all the resident threads per SM, GPU devices have large register files, with their size depending on the compute capability of the device; $8KB$ per SM for compute capability 1.0 and 1.1 devices, $16KB$ per SM for compute capability 1.2 and 1.3 GPUs, $32KB$ per SM for compute capability 2.0 devices and $64KB$ for compute capability 3.0 and 3.5 devices.

Shared memory is a small on-chip per SM memory space that has a low-latency access cost and can also be used to bypass the coalescing requirements of the global memory. Because it is shared between all threads of a thread block, it is usually used for synchronization as well as for data exchange between them. To enable concurrent accesses to it, shared memory is divided into $32$-bit memory banks. The maximum bandwidth of the shared memory is then utilized when a single memory request accesses one address from each different bank. If addresses from the same memory bank are accessed, the accesses are serialized to avoid conflicts with a significant degradation of the shared memory's bandwidth. Compute capability 1.0 to 1.3 GPUs have $16KB$ of shared memory per SM while compute capability 2.0 and newer GPUs have $48KB$ per SM.

Texture memory is a cached read-only memory space with a two-dimensional locality that is initialized host-side. Compute capability 1.x GPUs have an L1 texture cache per TPC and an L2 texture cache that is accessible by all SMs. It is often used to work around the coalescing requirements of the global memory and to increase the memory throughput of the kernel. The first time that an address of the texture memory is accessed, it is fetched from the global memory with the high latency that it entails. In that case though, the texture caches of the device are used to \textit{cache} the data, therefore minimizing the latency when cache hits occur. Unlike traditional CPU caches, the texture caches don't have coherency; writing data to texture memory either host- or device-side actually result in the invalidation of all caches.

The GPU has its own off-chip device memory, global memory. If data resides to \textit{pageable} host memory, a memory area that is usually allocated using \textit{malloc()}, it can be transferred to the GPU device explicitly before the launch of the kernel. Page-locked or \textit{pinned} memory, memory that always resides in physical memory as it cannot be paged out to disk, can also be allocated in host. Transferring of data between page-locked host memory and the global memory of the \textit{device} is performed concurrently with kernel execution using DMA to hide part of the latency involved. The global memory is accessible by all SMs using memory transactions of $32, 64$ and $128$ bytes with a high latency, usually between $400$ and $800$ clock cycles.

Local memory is a special type of memory that is a cacheable part of the global memory of the GPU. It is used when the registers of an SM are spilled. This can occur due to register pressure; when for example the execution context in terms of registers of a thread is higher than the hardware limit of the device. The term ``local'' refers to the fact that each thread has its own private area where its execution context is spilled, resolved at compilation time by the NVCC compiler.

\begin{figure}[h]
\centering
{\epsfig{file = 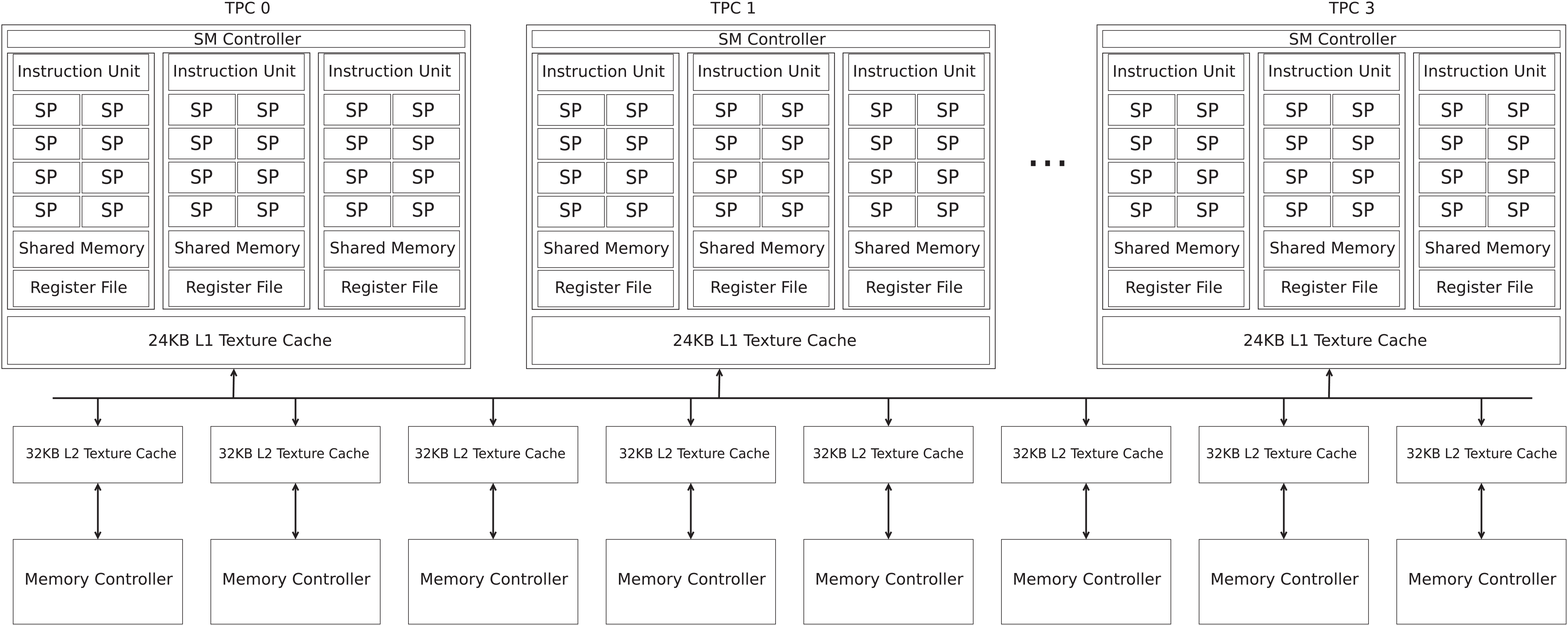, width = 12cm}}
  \caption{The NVIDIA GT200 Architecture}
  \label{fig:GT240}
\end{figure}

The NVIDIA GT 240 is a compute capability $1.2$ GPU from the GT200 series of NVIDIA's GeForce graphics processing units. It has $1GB$ of GDDR3 global memory, $550MHz$ Graphics clock rate, $1.34GHz$ Processor clock tester rate and $900MHz$ memory clock rate. As shown in Figure~\ref{fig:GT240}, it consists of $12$ SMs, with $3$ SMs per TPC. Each SM has $8$ SPs for a total of $96$ SPs. The SMs have $16KB$ of on-chip shared memory, and $16384$ $32$-bit registers. Each thread block can have a maximum of $512$ threads while each SM supports up to $1024$ active threads and up to $8$ active blocks. Each thread can use between $16$ registers when $1024$ threads are used per SM and $128$ as a maximum per-thread register usage. Since every SM contains one warp scheduler, one instruction is issued per warp over $4$ cycles. The latency of the arithmetic pipeline is $24$ cycles, therefore it can be completely hidden by having $6$ active warps at any one time. A request for any words within the same memory segment of the global memory for correctly aligned addresses is \textit{coalesced}, using one memory transaction per half-warp. The size of each global memory segment is $32$ bytes for $1$-byte words, $64$ bytes for $2$-byte words or $128$ bytes for words of $4, 8$ and $16$ bytes. The maximum memory throughput of the global memory can then be $128$ bytes per transaction. GTX 240 contains two levels of texture caches; a $24KB$ L1 cache within each TPC, partitioned in $3 \times 8KB$ caches and $8$ L2 texture caches with a size of $32KB$ each, visible to all SMs. The shared memory consists of $16$ banks organized in such a way that successive $32$-bit words are mapped into successive banks.\cite{CUDA_SDK} Each bank has a bandwidth of $32$ bits over two clock cycles and therefore the bandwidth of the shared memory is $64$ bytes over two cycles when all banks are accessed simultaneously. A request for shared memory addresses by a warp is split into two different requests, one for each half-warp.

\subsection{GPU Implementation}
\label{sec:GPU-Implementation}

The straightforward port of sequential applications to a GPU can often lead to significant speedups. The performance of the parallel implementations though can be improved even further when specific characteristics of the GPU architecture are taken into consideration. This section presents a basic data-parallel implementation strategy of the Aho-Corasick and Wu-Manber multiple pattern matching algorithms, analyzes the characteristics of the algorithm implementations that leverage the capabilities of the device, discusses the flaws that affect their performance and addresses them using different optimization techniques.
Table~\ref{tab:cuda_multi_notation} lists the notation that will be used for the rest of this section.

\begin{table}
\tbl{Implementation notation}
{\begin{tabular}{@{}ll@{}} \toprule
$numBlocks$ & The number of thread blocks \\
$blockDim$ & The size in threads of each block \\
$threadId$ & The unique ID of each thread of a block \\
$blockId$ & The unique ID of each block\\
$S_{chunk}$ & The size in characters of each chunk\\
$S_{thread}$ & The number of characters that each thread processes\\
$S_{memsize}$ & The size in bytes of the shared memory per thread block\\ \botrule
\end{tabular}}
\label{tab:cuda_multi_notation}
\end{table}

\subsubsection{Parallelization strategy}
\label{sec:parallelizationstrategy}

To expose the parallelism of the multiple pattern matching algorithms, the following basic data-parallel implementation strategy was used. The preprocessing phase of the algorithms was performed sequentially on the host CPU. The input string and all preprocessing arrays were copied to the global memory of the device. The input string was subsequently partitioned into $numBlocks$ character chunks, each with a size $S_{chunk}$ of $\frac{n}{numBlocks}$ characters. The chunks were then assigned to $numBlocks$ thread blocks. Each chunk was further partitioned into $blockDim$ sub-chunks, that in turn were assigned to each of the $blockDim$ threads of a thread block. To ensure the correctness of the results, $m - 1$ overlapping characters were used per thread. Therefore, each thread processed $S_{thread} = \frac{n}{numBlocks \times blockDim} + m - 1$ characters for a total of $(m - 1) ( numBlocks \times blockDim - 1 )$ additional characters. An integer array $out$ with a size of $numBlocks \times blockDim$ was used to store the number of matches per thread. To avoid extra coding complexity it is assumed that $n$ is divisible by both $numBlocks$ and $blockDim$. Since the character chunks have to overlap, the fewer possible thread blocks should be used to reduce the redundant characters as long as the maximum possible occupancy level is maintained per SM. 

Three tables were constructed during the preprocessing phase of the Aho-Corasick algorithm implementation. \textit{State\_transition} is a two-dimensional array where each row corresponds to a state of the trie, each column to a different character of the alphabet $\Sigma$ and the cells of the array represent the next state. To ensure that alignment requirements are met on each row, \textit{state\_transition} was allocated as pitched linear device memory using the \textit{cudaMallocPitch()} function. \textit{State\_supply} and \textit{state\_final} are one-dimensional arrays, allocated using the \textit{cudaMalloc()} function. Each column of the arrays corresponds to a different state of the trie while the cells represent the supply state of a given state and the information whether that state is final or not respectively. Since the Aho-Corasick trie can have a maximum of $m \times d + 1$ trie states, a value that for the experiments of this paper was typically equal to more than $65536$, each state was represented using a $4$-byte integer. Algorithm~\ref{compl:cuda_AC_basic_implementation} presents the basic data-parallel strategy used for the implementation of the Aho-Corasick algorithm using the CUDA API. As can be seen from the pseudocode, there can be divergence among the execution paths of the threads when previous states are visited using the \textit{supply} function of the algorithm.


\begin{algorithm}[h]

\textbf{Algorithm} Basic Aho-Corasick\\

$tid := blockId \times blockDim + threadId$\\
$start := ( blockId \times n ) / numBlocks + ( n \times threadId ) / ( numBlocks \times blockDim )$\\
$stop := start + n / ( numBlocks \times blockDim ) + m - 1$\\

$state := 0$\\

\For{ $i := start; i < stop; i := column + 1$ } {


		\While{ $newState := state\_transition[state, t_i] = fail$ }{
			
			$state := state\_supply[state]$\\
		}
		
		$state := newState$\\
	
		$out[tid] := out[tid] + state\_final[state]$\\
	
}

\caption{A basic parallel implementation of the Aho-Corasick algorithm}
\label{compl:cuda_AC_basic_implementation}
\end{algorithm}

The Wu-Manber algorithm uses the one-dimensional \textit{SHIFT} and the two-dimensional \textit{HASH} and \textit{PREFIX} tables. The space needed for the \textit{SHIFT} table is $\displaystyle\sum\limits_{i=0}^{B-1} |\Sigma| \times (2 ^ {bitshift})^i$. Each row of the \textit{HASH} and \textit{PREFIX} represents a different pattern from the pattern set while the columns represent different hash values. In the worst case there could be $d$ patterns with the same hash value $h$ or $h'$ for their $B$-character suffix or $B'$-character prefix respectively, therefore \textit{HASH} and \textit{PREFIX} require a $d \times \displaystyle\sum\limits_{i=0}^{B-1} |\Sigma| \times (2 ^ {bitshift})^i$ space each. The relevant data structures were copied to the global memory of the device with no modifications. The basic data-parallel strategy for the implementation of Wu-Manber is depicted in Algorithm~\ref{compl:cuda_WM_basic_implementation}.\\

\begin{algorithm}[h]

\textbf{Algorithm} Basic Wu-Manber\\

$id := blockId \times blockDim + threadId$\\
$start := ( blockId \times n ) / numBlocks + ( n \times threadId ) / ( numBlocks \times blockDim ) + m - 1$\\
$stop := start + n / ( numBlocks \times blockDim )$\\

$i := start$\\

\While{ $i < stop$} {

	$h := h_1(t_{i - B}\ldots t_i)$\\
	
	\uIf { $SHIFT[h] > 0$ } {
		$i := i + SHIFT[h]$\\
	}
	\Else {
		
		$h' := h_2(t_{i - m + 1}\ldots t_{i - m + 1 + B' - 1})$\\
		
		\ForAll{ pattern indices $r$ stored in $HASH[h]$ } {
		
			\If{ $PREFIX[r] = h'$ } {
		
				Verify the pattern corresponding to $r$ directly against the input string\\
			}
		}
		
		$i := i + 1$\\
	}
}

\caption{A basic parallel implementation of the Wu-Manber algorithm}
\label{compl:cuda_WM_basic_implementation}
\end{algorithm}

\subsubsection{Implementation limitations and optimization techniques}
\label{subsec:implementation-limitations-optimizations}

As detailed in section \ref{sec:GPU-Architecture}, accesses to global memory for compute capability $1.2$ GPUs by all threads of a half-warp are coalesced into a single memory transaction when all the requested words are within the same memory segment. The segment size is $32$ bytes when $1$-byte words are accessed, $64$ bytes for $2$-byte words and $128$ bytes for words of $4, 8$ and $16$ bytes. With the basic implementation strategy, each thread reads a single $1$-byte character on each iteration of the search loop; in this case the memory segment has a size of $32$ bytes. When $S_{thread} > 32$, each thread accesses a word from a different memory segment of the global memory. This results to uncoalesced memory transactions, with one memory transaction for every access of a thread. Since the maximum memory throughput of the global memory is $128$ bytes per transaction, the access pattern of the threads results in the utilization of only the $\frac{1}{128}$ of the available bandwidth.

To work around the coalescing requirements of the global memory and increase the utilization of the memory bandwidth, it is important to change the memory access pattern by reading words from the same memory segment and subsequently store them in the shared memory of the device. This involves the partition of the input string into $\frac{n}{S_{memsize}}$ chunks and the collective read of $S_{memsize}$ characters from the global into the shared memory by all $blockDim$ threads of a thread block. For each $16$ successive characters from the same segment then, only a single memory transaction will be used. This technique results in the utilization of the $\frac{1}{8}$ of the global memory bandwidth, improved by a factor of $16$. The threads can subsequently access the characters stored in shared memory in any order with a very low latency. Using the shared memory to increase the utilization of the memory bandwidth has two disadvantages. First, a total of $\frac{n}{S_{memsize}} \times ( blockDim - 1 ) \times ( m - 1 )$ redundant characters are used that introduce significantly more work overhead when compared to the basic data-parallel implementation strategy. Second, using the shared memory effectively reduces the occupancy of the SMs. As the size of the shared memory for each SM of the GTX 240 GPU is $16KB$, using the whole shared memory would reduce the occupancy to one thread block per SM. Partitioning the shared memory is not an efficient option since it would further increase the total work overhead.

The utilization of the global memory bandwidth can also increase when the threads read $16$-byte words instead of single characters on every memory transaction. For that, the built-in $uint4$ vector can be used, a C structure with members $x, y, z,$ and $w$ that is derived from the basic integer type. This way, each thread accesses an $128$-bit $uint4$ word that corresponds to $16$ characters of the input string with a single memory transaction while at the same time the memory segment size increases from $32$ to $128$ bytes. By having each thread read $128$-bit $uint4$ words from different memory segments results in the utilization of the $\frac{1}{8}$ of the global memory bandwidth similar to the coalescing technique above. The input string array stored in global memory can be casted to $uint4$ as follows:

\noindent\begin{minipage}{\textwidth}
\begin{lstlisting}
	
uint4 *uint4_text = reinterp_cast<uint4*>(d_text);

\end{lstlisting}
\end{minipage}\\

The two previous techniques can be combined; reading $16$ successive $128$-bit words or $256$ bytes in the form of $16$ $uint4$ vectors from global to shared memory can be done with just two memory transactions, fully utilizing the global memory bandwidth. The input string characters are then extracted from the uint4 vectors as retrieved from the global memory and are subsequently stored in shared memory on a character-by-character basis. To access the characters inside a uint4 vector, the vector can be recasted to $uchar4$:

\noindent\begin{minipage}{\textwidth}
\begin{lstlisting}	
uint4 uint4_var = uint4_text[i];

uchar4 uchar4_var0 = *reinterp_cast<uchar4*>(&uint4_var.x);
uchar4 uchar4_var1 = *reinterp_cast<uchar4*>(&uint4_var.y);
uchar4 uchar4_var2 = *reinterp_cast<uchar4*>(&uint4_var.z);
uchar4 uchar4_var3 = *reinterp_cast<uchar4*>(&uint4_var.w);
		
\end{lstlisting}
\end{minipage}\\

The drawback of casting input string characters to $uint4$ vectors and recasting them to $uchar4$ vectors is that it can be expensive in terms of processing power.

The preprocessing arrays of the algorithm are relatively small in size while at the same time they are frequently accessed by the threads. Moreover, as a character-by-character verification of the patterns to the input string is required, the pattern set array is also often accessed. The performance of the parallel implementation of the algorithm should then benefit from the binding of the relevant arrays to the texture memory of the device. The texture reference was bound to the device memory using \textit{cudaBindTexture()} for one-dimensional arrays and \textit{cudaBindTexture2D()} for two-dimensional arrays allocated as pitched linear device memory. The textures were then accessed in-kernel using the \textit{tex1Dfetch()} and \textit{tex2D()} functions. Arrays accessed via textures not only take advantage of the texture caches to minimize the memory latency when cache hits occur but also bypass the coalescing requirements of the global memory. Moreover, the maximum size for an one-dimensional texture reference is $8,192B$ while the maximum size for two-dimensional texture references is $65,536 \times 32,768$ \textit{texels}.

\clearpage

The shared memory of the GTX 240 GPU consists of $16$ memory banks numbered $0 - 15$. The banks are organized in such a way that successive $32$-bit words are mapped into successive banks with the $i^{th}$ word being stored in bank $i \bmod 16 - 1$. Bank conflicts occur when two or more threads of the same half-warp try to simultaneously access words $i, j,\ldots z$ when $i \bmod 16 = j \bmod 16 =\ldots = z \bmod 16$. When the memory coalescence optimizations described above are used, it is challenging to avoid bank conflicts when the $16$ characters of a $uint4$ vector are successively stored to the shared memory and when are retrieved from shared memory by the threads of the same half-warp during the search phase. Storing the input string characters in shared memory results in a $4$-way bank conflict. An alternative would be to cast each $uint4$ vector to $4$ $uchar4$ vectors and store them in shared memory in a round-robin fashion:

\noindent\begin{minipage}{\textwidth}
\begin{lstlisting}
int tid16 = threadIdx.x % 16;
				
if ( tid16 < 4 ) {
   uchar4_s_array[threadIdx.x * 4 + 0] = uchar4_var0;
   uchar4_s_array[threadIdx.x * 4 + 1] = uchar4_var1;
   uchar4_s_array[threadIdx.x * 4 + 2] = uchar4_var2;
   uchar4_s_array[threadIdx.x * 4 + 3] = uchar4_var3;
} else if ( tid16 < 8 ) {
   uchar4_s_array[threadIdx.x * 4 + 1] = uchar4_var1;
   uchar4_s_array[threadIdx.x * 4 + 2] = uchar4_var2;
   uchar4_s_array[threadIdx.x * 4 + 3] = uchar4_var3;
   uchar4_s_array[threadIdx.x * 4 + 0] = uchar4_var0;
} else if ( tid16 < 12 ) {
   uchar4_s_array[threadIdx.x * 4 + 2] = uchar4_var2;
   uchar4_s_array[threadIdx.x * 4 + 3] = uchar4_var3;
   uchar4_s_array[threadIdx.x * 4 + 0] = uchar4_var0;
   uchar4_s_array[threadIdx.x * 4 + 1] = uchar4_var1;
} else {
   uchar4_s_array[threadIdx.x * 4 + 3] = uchar4_var3;
   uchar4_s_array[threadIdx.x * 4 + 0] = uchar4_var0;
   uchar4_s_array[threadIdx.x * 4 + 1] = uchar4_var1;
   uchar4_s_array[threadIdx.x * 4 + 2] = uchar4_var2;
}		
\end{lstlisting}
\end{minipage}\\

This technique was not used since in practice the performance of the implementations did not improve. Although it was conflict-free when storing the vectors it resulted in a $4$-way thread divergence that serialized accesses to shared memory, the same effect that the example code was trying to avoid. The modulo operator is very expensive when used inside CUDA kernels and had a significant impact in the performance of the implementations. Algorithms~\ref{compl:cuda_AC_optimized_implementation} and~\ref{compl:cuda_WM_optimized_implementation} depict the pseudocode of the optimized kernel of the Aho-Corasick and Wu-Manber algorithms respectively. $sArray$ represents an array stored in the shared memory of the device with a size of $S_{memsize} = 16.128$ characters.

\begin{algorithm}[h]

\textbf{Algorithm} Optimized Aho-Corasick\\

$tid := blockId \times blockDim + threadId$\\
$start := S_{memsize} \times threadId / blockDim$\\
$stop := start + S_{memsize} / blockDim + m - 1$\\
$matches := 0$\\

\For { $gMem := blockId \times S_{memsize}; gMem < n; gMem := gMem + numBlocks \times S_{memsize}$ } {

	\For { $i := gMem / 16 + threadId, j := threadId; j < S_{memsize} / 16; i := i + blockDim, j := j + blockDim$ } {
		\If { $i < n / 16$ } {
			read a \textit{uint4} vector from $t_{i}$\\
			unpack the \textit{uint4} vector and store the $16$ characters to $sArray_{j * 16 + 0 \ldots j * 16 + 15}$\\
		}
	}
	Add $m - 1$ redundant characters to the end of the shared memory\\
	\textit{\_\_syncthreads()}\\
		
	$state := 0$\\
	
	\For { $i := start; i < stop$ \textbf{AND} $gMem + i < n; i := i + 1$ } {
		
			\While{ $newState := tex\_state\_transition[state, sArray_{i}] = fail$ }{
			
				$state := tex\_state\_supply[state]$\\
			}
	
			$matches := matches + tex\_state\_final[state]$\\
	}
	
	\textit{\_\_syncthreads()}\\
}
$out[tid] := matches$\\

\caption{An optimized parallel implementation of the Aho-Corasick algorithm}
\label{compl:cuda_AC_optimized_implementation}
\end{algorithm}

The $tex\_state\_transition$, $tex\_state\_supply$, $tex\_state\_final$, $tex\_SHIFT$, $tex\_HASH$ and $tex\_PREFIX$ arrays correspond to $state\_transition$, $state\_supply$, $state\_final$, $SHIFT$, $HASH$ and $PREFIX$ respectively when bound to the texture memory of the device. \\

\begin{algorithm}[h]

\textbf{Algorithm} Optimized Wu-Manber\\

$id := blockId \times blockDim + threadId$\\
$start := S_{memsize} \times threadId / blockDim + m - 1$\\
$stop := start + S_{memsize} / blockDim$\\
$matches := 0$\\

\For { $gMem := blockId \times S_{memsize}; gMem < n; gMem := gMem + numBlocks \times S_{memsize}$ } {

	\For { $i := gMem / 16 + threadId, j := threadIdx; j < S_{memsize} / 16; i := i + blockDim, j := j + blockDim$ } {
		\If { $i < n / 16$ } {
			read a \textit{uint4} vector from $t_{i}$\\
			unpack the \textit{uint4} vector and store the $16$ characters to $sArray_{j * 16 + 0 \ldots j * 16 + 15}$\\
		}
	}
	Add $m - 1$ redundant characters to the end of the shared memory\\
	\textit{synchronize threads}\\

	$i := start$\\

	\While{ $i < stop$} {

		$h := h_1(sArray_{i - B}\ldots sArray_i)$\\
	
		\uIf { $tex\_SHIFT[h] > 0$ } {
			$i := i + SHIFT[h]$\\
		}
		
		\Else {
	
			$h' := h_2(sArray_{i - m + 1}\ldots sArray_{i - m + 1 + B' - 1})$\\
		
			\ForAll{ pattern indices $r$ stored in $tex\_HASH[h]$ } {
		
				\If{ $tex\_PREFIX[r] = h'$ } {
		
					Verify the pattern corresponding to $r$ directly against the input string\\
				}
			}
		
			$i := i + 1$\\
		}
	}
}

$out[id] := matches$\\

\caption{An optimized parallel implementation of the Wu-Manber algorithm}
\label{compl:cuda_WM_optimized_implementation}
\end{algorithm}

\clearpage

\subsection{MPI Architecture}
Message Passing Interface (MPI) is a standardized and portable message-passing system, developed by a group of researchers from academia and industry to function on a wide variety of parallel computers. The first steps were made in the beginning of the nineties, and version $1.0$ of the interface was released in June 1994. MPI is considered the standard for High Performance Computing application development, on distributed memory architectures.\cite{Yang2011266}

MPI defines the semantics and the syntax of a core of library routines, useful to a wide range of developers writing portable message-passing programs. There are MPI bindings for many languages, including bindings for Fortran $77$, C and C++. There are several implementations of the MPI interface, while many of them are totally free and are available to the public domain.

One of the most widely used implementations of MPI is MPICH.\cite{mpich} It supports efficiently different computation and communication platforms including commodity clusters (desktop systems, shared-memory systems, multicore architectures), high-speed networks, proprietary high-end computing systems (Blue Gene, Cray) and multiple operating systems (Windows, most flavors of UNIX including Linux and MacOSX).

MPI is commonly used to implement parallel applications for cluster systems, as it handles the communication and the synchronization between the nodes. However, passing data over the network is a time consuming operation, therefore applications should balance the communication time with the processing time. In order to achieve this balance, the ratio of computation to communication time must be maintained as high as possible.

\subsection{MPI Implementation}
The nodes of the cluster, on which the experiments were executed, were connected through a Gigabit switch. The dataset was shared among the nodes using the Network File System\cite{Callaghan2000} protocol (NFS). Hence, once it was copied to the file system it was subsequently available to all nodes, without the need to explicitly distribute it using the scatter operation of MPI.

The input string was split into chunks, each of them assigned to a different node of the cluster. The split was achieved by using two auxiliary variables, \textit{start} and \textit{stop}, indicating the beginning and the end of each biological sequence chunk. The chunk assignment took place using the value of the \textit{MPI\_Rank} function that returns the identification number of each node, the \textit{MPI\_Size} function that returns the total number of nodes available, the size $n$ of the input string and the pattern size $m$. The two auxiliary variables were calculated using the following formulas:

\begin{equation}
\begin{array} {lllll} 
\displaystyle start & = & MPI\_Rank & * & \frac{n}{MPI\_Size}\\
\displaystyle stop & = (& MPI\_Rank+1) & * & \frac{n}{MPI\_Size} + (m-1)
\end{array}
\label{eq:chunk-assignments}
\end{equation}

After the memory allocation of the strings, the CPU and the GPU functions were initiated on each node, returning the result count of each search. Finally, the count of the multiple pattern occurrences was gathered from each worker node with the master node calculating the total sum using the \textit{MPI\_Reduce} function of MPI.

\section{Experimental Methodology}
\label{sec:experimental-methodology}

The performance of the parallel implementations of the algorithms was evaluated by comparing the running time of the GPU and Distributed GPU implementations to that of the sequential version. The algorithm implementations were processed on a computer cluster with $10$ homogeneous nodes, each with an Intel Core 2 Duo E8400 CPU, with two cores clocked at $3.00GHz$, $64KB$ L1 and $6MB$ L2 Cache, and $2GB$ of DDR2. Additionally, each node was equipped with a compute capability $1.2$ NVIDIA GT 240 GPU, with $1GB$ of GDDR3 global memory, $550MHz$ Graphics clock rate, $1.34GHz$ Processor clock tester rate and $900MHz$ memory clock rate. The cluster had a shared NFS disk space, while MPICH2 was used to handle all communication and synchronization operations, under Ubuntu Linux $12.04$. The algorithm implementations were compiled using MPICH2 and CUDA $5.5$, while the searching time of the sequential implementations was measured using the $MPI\_Wtime$ function of the Message Passing Interface since it has a better resolution than the standard $clock()$ function of \textit{time.h}. The searching time of the CUDA functions was measured using the CUDA event API.

For the experiments, a snapshot of the Expressed Sequence Tags (ESTs) from the February, 2009 assembly of the human genome was used, as produced by the Genome Reference Consortium and retrieved by the archives of the University of California Santa Cruz.\cite{ucsc2013} The snapshot, that was first converted to a one-dimensional input string and had any comments removed, consisted of $1,073,741,824$ characters and had an alphabet of $\Sigma=\{a, c, g, t\}$, the four nucleobases of the Deoxyribonucleic Acid (DNA).

The size of the data file was larger than the available size of \textit{global memory} of the GPU, thus it was impossible to load the input string on the device memory in one pass, and consequently to take advantage of the full computational power of the GPU. Host memory could be used, but by default, host memory allocations are pageable, while the GPU is unable to access data from pageable memory. Therefore, the data was allocated on pinned host memory and was then zero-copied to the GPU.

In order to simulate several demanding biological sequence searches, different multiple pattern sets were used. The sets were created from subsequences of the corresponding input string, consisting of $1,000$, $8,000$ and $16,000$ patterns, with each pattern having a size of $m = 8$ characters. The subsequences were chosen for at least $\min\{d, \lfloor \frac{n}{m}\rfloor\}$ matches.

Finally, the source code that was used to run the experiments is available at: https://github.com/iassael/hybrid\_cuda\_aho\_wu, under GNU General Public License.

\section{Experimental Results}
\label{sec:experimental-results}

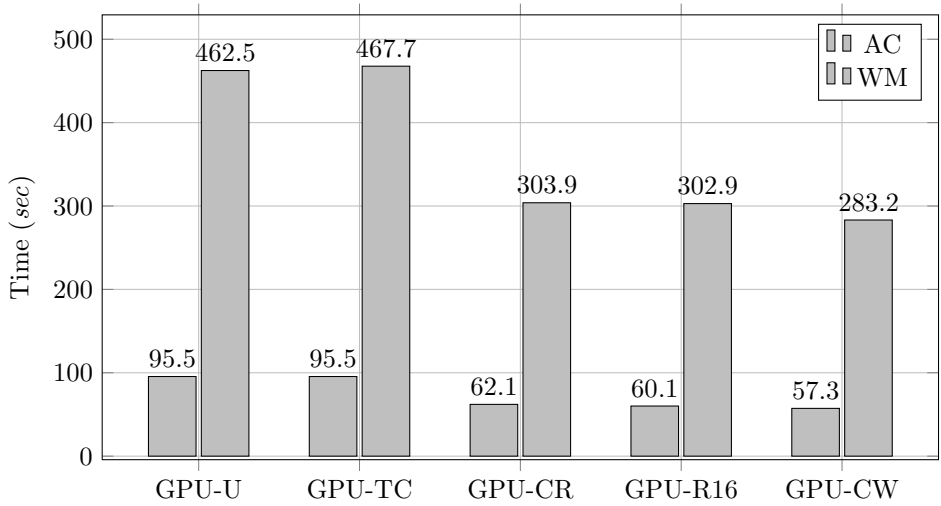
\begin{figure}[h]
\centering
\begin{tikzpicture}
	\begin{axis}[
		ybar,
		enlargelimits=0.15,		
		bar width=18pt,
		height=7.5cm,
		width=\textwidth,
		symbolic x coords={GPU-U, GPU-TC, GPU-CR, GPU-R16, GPU-CW},
		ylabel=Time (\textit{sec}),
              grid=both,
              nodes near coords,
		xtick=data
	]
	\addplot  +[
	black,
	fill=lightgray,
	] coordinates {
		(GPU-U,  95.5)
		(GPU-TC,  95.5)
		(GPU-CR,  62.1)
		(GPU-R16,  60.1)
		(GPU-CW,  57.3)
	};
	\addplot  +[
	black,
	fill=lightgray,
	postaction={
		pattern=north east lines
	}
	] coordinates {
		(GPU-U,  462.5)
		(GPU-TC,  467.7)
		(GPU-CR,  303.9)
		(GPU-R16,  302.9)
		(GPU-CW,  283.2)
	};
	\legend{AC\\WM\\}
	\end{axis}
\end{tikzpicture}
\caption{Aho-Corasick \& Wu-Manber GPU Optimizations}
\label{fig:acwm_GPU-optimizations}
\end{figure}

This section evaluates the performance of the Aho-Corasick and Wu-Manber algorithm implementations. As stated in section \ref{subsec:implementation-limitations-optimizations}, different optimization techniques were used for the implementation of the algorithms in order to take full advantage of the device hardware. Figure \ref{fig:acwm_GPU-optimizations} illustrates the execution times of each of the five GPU optimizations, running on a single node. For the execution, a set of $8,000$ patterns was used, each with a pattern size of $m=8$. Each implementation stage also incorporates the optimizations of the previous stages.

\begin{enumerate}
	
	\item The first stage of the implementation was unoptimized. The input string, the pattern set, the preprocessing arrays of the algorithms and the $out$ array that holds the number of matches per thread, were stored in the global memory of the device. Each thread accessed input string characters directly from the global memory as needed and stored the number of matches directly to $out$. At the end of the algorithms' search phase, the results array was transferred to the host memory and the total number of matches was calculated by the CPU. (GPU-U)
	
	\item The second stage of the implementation involved the binding of the preprocessing arrays and the pattern set array in the case of the Wu-Manber algorithm to the texture memory of the device. This optimization was not applied on the Aho-Corasick implementation as the preprocessing tables take more space than the supplied GPU texture memory for large pattern sets. (GPU-TC)
	
	\item In the third stage, the threads worked around the coalescing requirements of the global memory by collectively reading input string characters and storing them to the shared memory of the device. (GPU-CR)
	
	\item The fourth stage of the implementation was similar to the third but in addition, each thread read simultaneously $16$ input string characters from the global memory by using a $uint4$ vector. The characters were extracted from the vectors using $uchar4$ vectors and were subsequently stored to the shared memory. (GPU-R16)
	
	\item The fifth implementation stage involves the coalescing of the writes by the threads to the global memory of the device. (GPU-CW)
	
\end{enumerate}

As depicted in Figure \ref{fig:acwm_GPU-optimizations}, the final optimized parallel implementation of the Aho-Corasick and the Wu-Manber algorithms was $1.66$x and $1.63$x faster than their unoptimized version respectively. This indicates the significant performance increase that can be achieved when the implementations are customized to take advantage of the specific underlying hardware.

\begin{figure}[h]
\centering
\begin{tikzpicture}
        \begin{axis}
	[
		height=7.5cm,
		width=\textwidth,
		grid=both,		
		xlabel=Number of processing nodes,
        	ylabel=Time (\textit{sec})
	]
	\addplot [color = gray, mark=triangle*] coordinates{
		(1, 26.08110621 ) (2, 13.0635242 ) (3, 9.679050247 ) (4, 6.522398536 ) (5, 5.289730038 ) (6, 4.375002936 ) (7, 3.768002724 ) (8, 3.285003587 ) (9, 3.078089243 ) (10, 2.637820736)
	};
	\addplot [color = black, mark=square*]  coordinates{
		(1, 38.795927 ) (2, 18.427936 ) (3, 12.827447 ) (4, 9.776997 ) (5, 7.893712 ) (6, 6.660501 ) (7, 5.776665 ) (8, 5.102328 ) (9, 4.61638 ) (10, 4.17171)
	};
	\addplot [color = darkgray, mark=diamond*]  coordinates{
		( 1 , 17.448042 )
( 2 , 12.46201 )
( 3 , 12.76181 )
( 4 , 11.3498 )
( 5 , 11.564684 )
( 6 , 10.714231 )
( 7 , 12.213895 )
( 8 , 11.790964 )
( 9 , 12.360423 )
( 10 , 14.63768 )
	};
            
	\legend{AC 1000 patterns\\WM 1000 patterns\\Network Read \& Gather\\}
        \end{axis}
\end{tikzpicture}
\caption{Aho-Corasick \& Wu-Manber Cluster's Nodes Used Comparison for $1,000$ patterns}
\label{fig:acwm_nodes1000}
\end{figure}
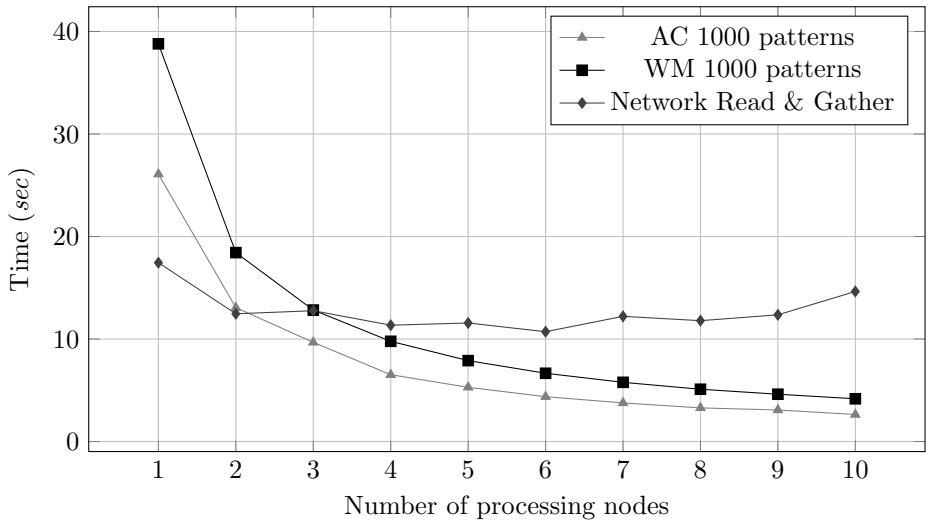

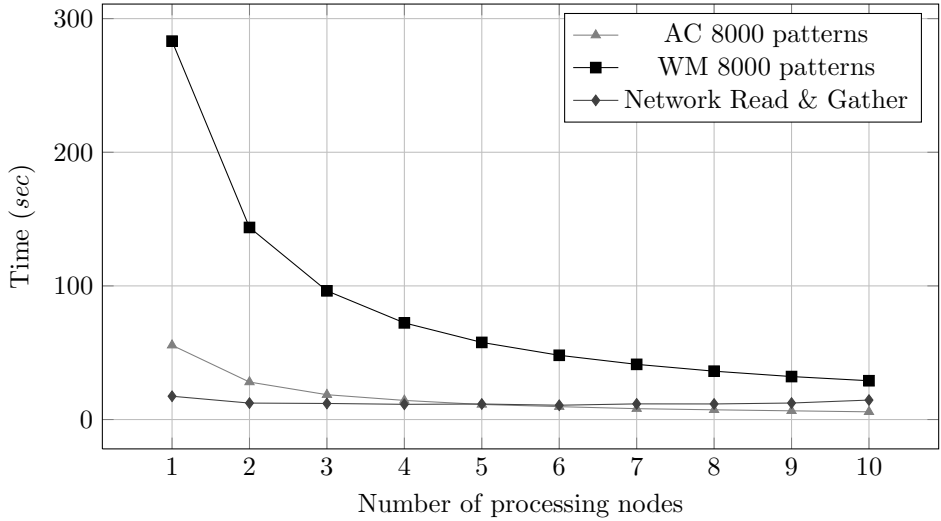
\begin{figure}[h]
\centering
\begin{tikzpicture}
        \begin{axis}
	[
		height=7.5cm,
		width=\textwidth,
		grid=both,		
		xlabel=Number of processing nodes,
        	ylabel=Time (\textit{sec})
	]
	\addplot [color = gray, mark=triangle*] coordinates{
		( 1 , 55.59044781 ) ( 2 , 28.0574279 ) ( 3 , 18.58200313 ) ( 4 , 14.30901747 ) ( 5 , 11.27856157 ) ( 6 , 9.640428827 ) ( 7 , 8.130614679 ) ( 8 , 7.275712177 ) ( 9 , 6.515073469 ) ( 10 , 5.771265665 )
	};
	\addplot [color = black, mark=square*]  coordinates{
		(1, 283.156952 ) (2, 143.72614 ) (3, 96.301426 ) (4, 72.324056 ) (5, 57.709293 ) (6, 48.073995 ) (7, 41.292983 ) (8, 36.183308 ) (9, 32.162771 ) (10, 29.039478)
	};
	\addplot [color = darkgray, mark=diamond*]  coordinates{
		( 1 , 17.430959 )
( 2 , 12.308856 )
( 3 , 12.026645 )
( 4 , 11.384426 )
( 5 , 11.614029 )
( 6 , 10.704602 )
( 7 , 11.717291 )
( 8 , 11.666787 )
( 9 , 12.352483 )
( 10 , 14.565721 )
	};
            
	\legend{AC 8000 patterns\\WM 8000 patterns\\Network Read \& Gather\\}
        \end{axis}
\end{tikzpicture}
\caption{Aho-Corasick \& Wu-Manber Cluster's Nodes Used Comparison for $8,000$ patterns}
\label{fig:acwm_nodes8000}
\end{figure}

\begin{figure}[h!]
\centering
\begin{tikzpicture}
        \begin{axis}
	[
		height=7.5cm,
		width=\textwidth,
		grid=both,		
		xlabel=Number of processing nodes,
        	ylabel=Time (\textit{sec})
	]
	\addplot [color = gray, mark=triangle*] coordinates{
		( 1 , 66.69782057 ) ( 2 , 33.73798385 ) ( 3 , 22.26686065 ) ( 4 , 17.38175948 ) ( 5 , 13.61955926 ) ( 6 , 11.73746916 ) ( 7 , 9.90550025 ) ( 8 , 8.95157835 ) ( 9 , 8.028906633 ) ( 10 , 7.107128268 )
	};
	\addplot [color = black, mark=square*]  coordinates{
		(1, 546.140773 ) (2, 275.746245 ) (3, 183.719287 ) (4, 138.484725 ) (5, 111.464913 ) (6, 91.972058 ) (7, 79.129625 ) (8, 69.206201 ) (9, 61.526889 ) (10, 55.257791)
	};
	\addplot [color = darkgray, mark=diamond*]  coordinates{
		( 1 , 17.422472 )
( 2 , 12.305245 )
( 3 , 12.026315 )
( 4 , 11.364924 )
( 5 , 11.53139 )
( 6 , 10.678401 )
( 7 , 11.700588 )
( 8 , 11.674269 )
( 9 , 12.306429 )
( 10 , 14.513983 )
	};
            
	\legend{AC 16000 patterns\\WM 16000 patterns\\Network Read \& Gather\\}
        \end{axis}
\end{tikzpicture}
\caption{Aho-Corasick \& Wu-Manber Cluster's Nodes Used Comparison for $16,000$ patterns}
\label{fig:acwm_nodes16000}
\end{figure}
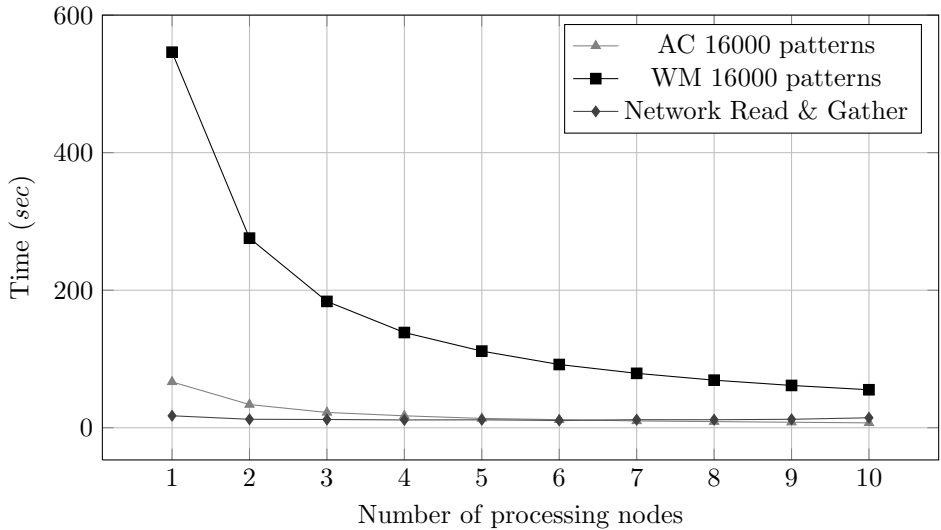

Figures \ref{fig:acwm_nodes1000} to \ref{fig:acwm_nodes16000} present the execution times of the searching phase of the parallel implementations for both the Aho-Corasick and Wu-Manber algorithms, including the time to distribute the data to the worker nodes of the cluster using the NFS protocol and the time to gather the results back to the master node using MPI. The pattern sets that were used, consisted of $1,000$, $8,000$ and $16,000$ patterns while at the same time, $1$ to $10$ cluster nodes were used.

As presented in Figures \ref{fig:acwm_nodes1000} to \ref{fig:acwm_nodes16000}, parameters such as the number of the nodes used and the pattern set size affect significantly the execution time of the algorithm implementations. As the number of worker nodes increased, the total network communication time also increased. Moreover, for sets of $1,000$ patterns, the network communication time was a large percentage of the total execution time. As can be observed in Figure \ref{fig:acwm_nodes1000}, this time overlaps the parallel searching time when using more than $2$ nodes for Aho-Corasick and $4$ or more nodes for Wu-Manber. Even for larger pattern sets, the network communication time is a significant proportion compared to the execution times of the implementations, especially in the case of the Aho-Corasick algorithm. As it can be seen in Figures \ref{fig:acwm_nodes8000} and \ref{fig:acwm_nodes16000}, the search execution time is approximately equal to the network communication time when using more than $5$ nodes. On the other hand and for larger pattern sets, it consists a smaller proportion of the total execution time of the Wu-Manber algorithm, where the searching phase of the algorithm implementation is much more demanding.

It is noteworthy that for all the experiments, Aho-Corasick outperformed the Wu-Manber algorithm. For smaller pattern sets the difference was not significant, while for larger sets of patterns this difference became substantial. More specifically, for pattern sets of $1,000$, $8,000$ and $16,000$, Aho-Corasick was approximately $1.49x$, $5.05x$ and $7.97x$ times faster than the Wu-Manber algorithm respectively.

\begin{table}[h]
\tbl{Comparison of execution times for $1$ to $10$ cluster nodes and for sets of $1,000$ patterns }
{\begin{tabular}{@{}c |ccccc@{}} \toprule
Nodes & NFS & Load Files & AC GPU & WM GPU & Reduce\\
\colrule
1 & 5.04 & 12.40 & 26.08 & 38.80 & 0.00\\
2 & 6.10 & 6.36 & 13.06 & 18.43 & 0.00\\
3 & 7.85 & 4.91 & 9.68 & 12.83 & 0.00\\
4 & 7.51 & 3.13 & 6.52 & 9.78 & 0.71\\
5 & 7.07 & 2.50 & 5.29 & 7.89 & 2.00\\
6 & 6.09 & 2.14 & 4.38 & 6.66 & 2.49\\
7 & 6.99 & 2.31 & 3.77 & 5.78 & 2.91\\
8 & 6.96 & 1.68 & 3.29 & 5.10 & 3.15\\
9 & 7.56 & 1.45 & 3.08 & 4.62 & 3.36\\
10 & 8.91 & 1.37 & 2.64 & 4.17 & 4.35\\
\hline
\end{tabular}}
\begin{tabnote}
The displayed time values are in \textit{sec}, pattern size m=$8$
\end{tabnote}
\label{tbl:comp-execution-1000}
\end{table}

Detailed average execution times of each part of the implemented algorithms, running on sets of $1,000$, $8,000$, $16,000$ patterns with a pattern size of $m=8$, are given in Tables \ref{tbl:comp-execution-1000}, \ref{tbl:comp-execution-8000} and \ref{tbl:comp-execution-16000} respectively. These tables illustrate the difference between the proportion of the NFS file distribution time, the time to load the data and the time to reduce the results back to the master node using the MPI\_Reduce function, and the corresponding time to execute the searching phase of an algorithm for each different pattern set.

\begin{table}[h]
\tbl{Comparison of execution times for $1$ to $10$ cluster nodes and for sets of $8,000$ patterns}
{\begin{tabular}{@{}c |ccccc@{}} \toprule
Nodes & NFS & Load Files & AC GPU & WM GPU & Reduce\\
\colrule
1 & 5.04 & 12.40 & 55.59 & 300.60 & 0.00\\
2 & 6.10 & 6.36 & 28.06 & 156.19 & 0.00\\
3 & 7.85 & 4.91 & 18.58 & 109.06 & 0.00\\
4 & 7.51 & 3.13 & 14.31 & 83.67 & 0.71\\
5 & 7.07 & 2.50 & 11.28 & 69.27 & 2.00\\
6 & 6.09 & 2.14 & 9.64 & 58.79 & 2.49\\
7 & 6.99 & 2.31 & 8.13 & 53.51 & 2.91\\
8 & 6.96 & 1.68 & 7.28 & 47.97 & 3.15\\
9 & 7.56 & 1.45 & 6.52 & 44.52 & 3.36\\
10 & 8.91 & 1.37 & 5.77 & 43.68 & 4.35\\
\hline
\end{tabular}}
\begin{tabnote}
The displayed time values are in \textit{sec}, pattern size m=$8$
\end{tabnote}
\label{tbl:comp-execution-8000}
\end{table}

\begin{table}[h]
\tbl{Comparison of execution times for $1$ to $10$ cluster nodes and for sets of $16,000$ patterns}
{\begin{tabular}{@{}c |ccccc@{}} \toprule
Nodes & NFS & Load Files & AC GPU & WM GPU & Reduce\\
\colrule
1 & 5.04 & 12.40 & 66.70 & 546.14 & 0.00\\
2 & 6.10 & 6.36 & 33.74 & 275.75 & 0.00\\
3 & 7.85 & 4.91 & 22.27 & 183.72 & 0.00\\
4 & 7.51 & 3.13 & 17.38 & 138.48 & 0.71\\
5 & 7.07 & 2.50 & 13.62 & 111.46 & 2.00\\
6 & 6.09 & 2.14 & 11.74 & 91.97 & 2.49\\
7 & 6.99 & 2.31 & 9.91 & 79.13 & 2.91\\
8 & 6.96 & 1.68 & 8.95 & 69.21 & 3.15\\
9 & 7.56 & 1.45 & 8.03 & 61.53 & 3.36\\
10 & 8.91 & 1.37 & 7.11 & 55.26 & 4.35\\
\hline
\end{tabular}}
\begin{tabnote}
Pattern size m=$8$
\end{tabnote}
\label{tbl:comp-execution-16000}
\end{table}

The overall speedup of each of the GPU implementations, under the different pattern set sizes and running on multiple nodes is presented in Table \ref{tbl:overall-execution-speedup}. The speedup is calculated by the total execution time of the Aho-Corasick and Wu-Manber algorithms, as depicted in Tables \ref{tbl:comp-execution-1000} to \ref{tbl:comp-execution-16000}, compared to the corresponding single-node performance. It is worth noting, that for a pattern set size of $1,000$ patterns, both algorithms exhibited a significant speedup with an upward trend when up to $5$ nodes were used. For more than $5$ nodes, the time of MPI\_Reduce had the tendency to increase proportionally to the number of nodes, affecting significantly the total execution time. Both algorithms exhibited a maximum speedup of $2.89x$ and $3.33x$ respectively, when they were executed on a cluster of $8$ nodes. A similar trend was observed for pattern sets of $8,000$, where the maximum speedup for both algorithms was exhibited using $9$ cluster nodes, and was equal to $3.87x$ and $5.59x$ for AC and WM respectively. Finally, for pattern sets of $16,000$, the speedup of the WM algorithm had a gradual increase by the number of cluster nodes, levelling at $8.06x$ when the algorithm was executed on all the available nodes. On the other hand, the AC algorithm had a maximum speedup of $4.13x$ when it was executed on $9$ cluster nodes.

 
\begin{table}[h]
\tbl{Overall speedup for sets of $1,000$, $8,000$ and $16,000$ patterns}
{\begin{tabular}{@{}c | cc |cc|cc@{}} \toprule
Nodes & AC 1000 & WM 1000 & AC 8000 & WM 8000 & AC 16000 & WM 16000\\
\colrule
1 & 1.00x & 1.00x & 1.00x & 1.00x & 1.00x & 1.00x\\
2 & 1.71x & 1.82x & 1.80x & 1.89x & 1.82x & 1.96x\\
3 & 1.94x & 2.20x & 2.33x & 2.61x & 2.40x & 2.87x\\
4 & 2.44x & 2.66x & 2.85x & 3.35x & 2.93x & 3.76x\\
5 & 2.58x & 2.89x & 3.20x & 3.93x & 3.34x & 4.58x\\
6 & 2.88x & 3.24x & 3.59x & 4.58x & 3.75x & 5.49x\\
7 & 2.72x & 3.13x & 3.59x & 4.84x & 3.80x & 6.17x\\
8 & 2.89x & 3.33x & 3.83x & 5.32x & 4.06x & 6.96x\\
9 & 2.82x & 3.31x & 3.87x & 5.59x & 4.13x & 7.63x\\
10 & 2.52x & 2.99x & 3.58x & 5.45x & 3.87x & 8.06x\\
\hline
\end{tabular}}
\begin{tabnote}
The displayed time values are in \textit{sec}, pattern size m=$8$
\end{tabnote}
\label{tbl:overall-execution-speedup}
\end{table}


\clearpage

\section{Conclusions}
\label{sec:conclusions}

This paper presented parallel implementations of two of the most well known multiple matching algorithms, Aho-Corasick and Wu-Manber, on a homogeneous cluster of nodes using the MPI and CUDA APIs. The performance of the algorithm implementations was evaluated when executed on an subsequence of the Expressed Sequence Tags of the human genome for different number of cluster nodes and pattern set sizes. The algorithm implementations were optimized in different steps in order to take advantage of the underlying GPU hardware. It was generally discussed that even low-end GPU cards could considerably facilitate demanding tasks, such as the processing of biological sequences. Finally, it was proven that a substantial performance increase can be achieved, when taking advantage of the combined power of the GPU nodes of a cluster.

The performance of the parallel algorithm implementations was evaluated over the corresponding time running on a single computer node, and was compared to the distributed implementation when executed on a GPU cluster of $10$ nodes. 

Based on the results, it was determined that the searching time of the parallel algorithm implementations was between $9.38x$ and $9.88x$ over the sequential version for the Aho-Corasick algorithm, and between $9.29x$ and $9.88x$ for the Wu-Manber algorithm. It was also determined that the overall speedup for each algorithm was affected by the network communication time, proportionately to the number of nodes used. More specifically and for sets of $1,000$ patterns, the overall speedup for both the Aho-Corasick and the Wu-Manber algorithm implementations increased when up to $5$ cluster nodes were used, leveling at $2.88x$ and $3.24x$ respectively, with the maximum speedups of $2.89x$ and $3.33x$ being exhibited for $8$ cluster nodes. For sets of $8,000$ patterns, the maximum measured speedup was $3.87x$ for the Aho-Corasick algorithm implementation using $9$ nodes, and $6.88x$ for the Wu-Manber algorithm implementation making use of all the available nodes of the cluster. Finally, for sets of $16,000$ patterns, the equivalent speedups were $4.13x$ for Aho-Corasick using $9$ nodes, and $8.06x$ for Wu-Manber using $10$ nodes.

Although the Wu-Manber algorithm implementation exhibited a significant speedup in all cases, it was outperformed by the Aho-Corasick algorithm implementation for a pattern size of $m=8$, especially in larger pattern sets. For a sequential execution of the algorithm implementations and for sets of $1,000$, $8,000$ and $16,000$ patterns, Aho-Corasick was $1.29x$, $4.11x$ and $6.69x$ faster than Wu-Manber, and therefore it can be considered as a better choice for searching Biological Sequence Databases.

\bibliographystyle{ws-procs11x85}
\bibliography{cudampiwm}

\begin{thebibliography}{10}

\bibitem{gpgpuorg}
GPGPU, {General-Purpose computation on Graphics Processing Units} Website,
  http://gpgpu.org/.

\bibitem{CUDA_SDK}
NVIDIA, {NVIDIA CUDA Compute Unified Device Architecture Programming Guide,
  version 5.5}  (2013).

\bibitem{Kouzinopoulos2011}
C.~Kouzinopoulos, P.~Michailidis and K.~Margaritis, {\em Parallel Processing of
  Multiple Pattern Matching Algorithms for Biological Sequences: Methods and
  Performance Results} (InTech, 2011).

\bibitem{Aho1975}
A.~Aho and M.~Corasick, {Efficient String Matching: An Aid to Bibliographic
  Search}, {\em Communications of the ACM} {\bf 18}, 333  (1975).

\bibitem{Wu1994}
S.~Wu and U.~Manber, {\em {A Fast Algorithm for Multi-pattern Searching}},
  tech. rep.  (1994), Technical report TR-94-17.

\bibitem{Huang2008}
N.-F. Huang, H.-W. Hung, S.-H. Lai, Y.-M. Chu and W.-Y. Tsai, {A GPU-Based
  Multiple-Pattern Matching Algorithm for Network Intrusion Detection Systems},
  in {\em Proceedings of the 22nd International Conference on Advanced
  Information Networking and Applications - Workshops\/}, AINAW '08 (IEEE
  Computer Society, Washington, DC, USA, 2008).

\bibitem{SnortWeb}
Snort, {Snort Intrusion Prevention and Detection System} Website,
  http://www.snort.org/.

\bibitem{Hongjian2011}
H.~Li, B.~Ni, M.-H. Wong and K.-S. Leung, {A Fast CUDA Implementation of Agrep
  Algorithm for Approximate Nucleotide Sequence Matching}, in {\em Application
  Specific Processors (SASP), 2011 IEEE 9th Symposium on\/}, 2011.

\bibitem{Wu1992}
S.~Wu and U.~Manber, {Agrep - A Fast Approximate Pattern-Matching Tool}, {\em
  In Proceedings of USENIX Technical Conference} , 153  (1992).

\bibitem{tran2011}
T.~T. Tran, M.~Giraud and J.-S. Varr{\'e}, {Bit-Parallel Multiple Pattern
  Matching}, in {\em {Parallel Processing and Applied Mathematics / Parallel
  Biocomputing Conference (PPAM / PBC 11)}\/},  (Torun, Pologne, 2011).

\bibitem{pyrgiotis2012parallel}
T.~Pyrgiotis, C.~Kouzinopoulos and K.~Margaritis, {\em Parallel Implementation
  of the Wu-Manber Algorithm Using the OpenCL Framework}, IFIP Advances in
  Information and Communication Technology, Vol.~382 (Springer Berlin
  Heidelberg, 2012).

\bibitem{Kouzinopoulos2012}
C.~S. Kouzinopoulos, P.~D. Michailidis and K.~G. Margaritis, {Performance Study
  of Parallel Hybrid Multiple Pattern Matching Algorithms for Biological
  Sequences}, in {\em BIOINFORMATICS\/}, 2012.

\bibitem{Vasiliadis2008}
G.~Vasiliadis, S.~Antonatos, M.~Polychronakis, E.~Markatos and S.~Ioannidis,
  {Gnort: High Performance Network Intrusion Detection Using Graphics
  Processors}, {\em Proceedings of RAID} {\bf 5230}, 116  (2008).

\bibitem{Vasiliadis2010}
G.~Vasiliadis and S.~Ioannidis, {Gravity: A Massively Parallel Antivirus
  Engine}, in {\em Recent Advances in Intrusion Detection\/}, 2010.

\bibitem{Lin2010}
C.~Lin, S.~Tsai, C.~Liu, S.~Chang and J.~Shyu, {Accelerating String Matching
  Using Multi-Threaded Algorithm on GPU}, in {\em Global Telecommunications
  Conference (GLOBECOM 2010), 2010 IEEE\/}, 2010.

\bibitem{Tumeo2011}
A.~Tumeo, S.~Secchi and O.~Villa, {Experiences with String Matching on the
  Fermi Architecture}, {\em Architecture of Computing Systems-ARCS 2011} , 26
  (2011).

\bibitem{Zha2011}
X.~Zha and S.~Sahni, {Multipattern String Matching on a GPU}, in {\em Computers
  and Communications (ISCC), 2011 IEEE Symposium on\/}, 2011.

\bibitem{Hu2012}
L.~Hu, Z.~Wei, F.~Wang, X.~Zhang and K.~Zhao, {An Efficient AC Algorithm with
  GPU}, {\em Procedia Engineering} {\bf 29}, 4249  (2012).

\bibitem{Jamshed2012}
M.~Jamshed, J.~Lee, S.~Moon, I.~Yun, D.~Kim, S.~Lee, Y.~Yi and K.~Park,
  {Kargus: a Highly-scalable Software-based Intrusion Detection System}, in
  {\em Proceedings of the ACM Conference on Computer and Communications
  Security (CCS)\/}, 2012.

\bibitem{Pungila2012}
C.~Pungila and V.~Negru, {A Highly-Efficient Memory-Compression Approach for
  GPU-Accelerated Virus Signature Matching}, {\em Information Security} , 354
  (2012).

\bibitem{Tumeo2012}
A.~Tumeo, O.~Villa and D.~Chavarr{\'\i}a-Miranda, {Aho-Corasick String Matching
  on Shared and Distributed-Memory Parallel Architectures}, {\em Parallel and
  Distributed Systems, IEEE Transactions on} {\bf 23}, 436  (2012).

\bibitem{Navarro2002}
G.~Navarro and M.~Raffinot, {\em {Flexible pattern matching in strings:
  practical on-line search algorithms for texts and biological sequences}}
  (Cambridge University Press, 2002).

\bibitem{Dori2006}
S.~Dori and G.~Landau, {Construction of Aho Corasick Automaton in Linear Time
  for Integer Alphabets}, {\em Information Processing Letters} {\bf 98}, 66
  (2006).

\bibitem{WEB03}
Streamline, {The Official Webpage of the Streamline Project} Website,  (2011),
  http://netstreamline.org/.

\bibitem{Chen2005}
X.~Chen, B.~Fang, L.~Li and Y.~Jiang, {WM+: An Optimal Multi-pattern String
  Matching Algorithm Based on the WM Algorithm}, {\em Advanced Parallel
  Processing Technologies} , 515  (2005).

\bibitem{Navarro2004}
G.~Navarro and K.~Fredriksson, {Average Complexity of Exact and Approximate
  Multiple String Matching}, {\em Theoretical Computer Science} {\bf 321}, 283
  (2004).

\bibitem{Liu2005}
P.~Liu, Y.~Liu and J.~Tan, A partition-based efficient algorithm for large
  scale multiple-strings matching, in {\em SPIRE\/}, 2005.

\bibitem{Wilt2013}
N.~Wilt, {\em {The CUDA Handbook: A Comprehensive Guide to GPU Programming}}
  (Addison-Wesley Professional, 2013).

\bibitem{Halfhill2008}
T.~Halfhill, {Parallel Processing with CUDA}, {\em Microprocessor Report} , 01
  (2008).

\bibitem{Lakshminarayana2010}
N.~Lakshminarayana and H.~Kim, {Effect of Instruction Fetch and Memory
  Scheduling on GPU Performance}, in {\em Workshop on Language, Compiler, and
  Architecture Support for GPGPU\/}, 2010.

\bibitem{Papadopoulou2009}
M.~Papadopoulou, M.~Sadooghi-Alvandi and H.~Wong, {Micro-benchmarking the GT200
  GPU}, {\em Computer Group, ECE, University of Toronto}   (2009), Technical
  report.

\bibitem{Volkov2010}
V.~Volkov, {Better Performance at Lower Occupancy}, in {\em Proceedings of the
  GPU Technology Conference, GTC\/}, 2010.

\bibitem{Gebhart2012}
M.~Gebhart, S.~Keckler, B.~Khailany, R.~Krashinsky and W.~Dally, {Unifying
  Primary Cache, Scratch, and Register File Memories in a Throughput
  Processor}, in {\em Proceedings of the 45th Annual IEEE/ACM International
  Symposium on Microarchitecture\/}, 2012.

\bibitem{Yang2011266}
C.-T. Yang, C.-L. Huang and C.-F. Lin, {Hybrid CUDA, OpenMP, and MPI Parallel
  Programming on Multicore GPU Clusters }, {\em Computer Physics
  Communications} {\bf 182}, 266   (2011).

\bibitem{mpich}
{MPICH} {\textbar} high-performance portable {MPI} Website,
  http://www.mpich.org/.

\bibitem{Callaghan2000}
B.~Callaghan, {\em {NFS Illustrated}} (Addison-Wesley Longman Ltd., Essex, UK,
  UK, 2000).

\bibitem{ucsc2013}
U.~of~California Santa~Cruz, Jack baskin school of engineering Website,
  (2013), http://hgdownload.cse.ucsc.edu/goldenPath/hg19/bigZips/.

\end{thebibliography}



\end{document}